\documentclass[fleqn,usenatbib]{mnras}

\usepackage{newtxtext,newtxmath}

\usepackage[T1]{fontenc}

\DeclareRobustCommand{\VAN}[3]{#2}
\let\VANthebibliography\thebibliography
\def\thebibliography{\DeclareRobustCommand{\VAN}[3]{##3}\VANthebibliography}

\usepackage{graphicx}	
\usepackage{amsmath}	
\usepackage{tabularx}
\usepackage{soul}

\newcommand{\RNum}[1]{\uppercase\expandafter{\romannumeral #1\relax}}
\DeclareMathOperator{\sech}{sech}

\title[Building stellar bulges and halo cores]{Building stellar bulges and halo cores from massive clumps observed in the DYNAMO-HST sample}

\author[Mahmoud Hashim]{
Mahmoud Hashim,$^{1}$\thanks{mahmoud.hashim@bue.edu.eg}
Amr A. El-Zant,$^{1}$
\thanks{amr.elzant@bue.edu.eg}
and Antonino Del Popolo$^{2,3,4}$
\\
$^{1}$Centre for Theoretical Physics, The British University in Egypt, Sherouk City, 11837 Cairo, Egypt\\
$^{2}$Dipartimento di Fisica e Astronomia, University Of Catania, Viale Andrea Doria 6, 95125, Catania, Italy\\
$^{3}$ Institute of Astronomy and National Astronomical Observatory, Bulgarian Academy of Sciences, 72, Tsarigradsko Shosse Blvd., 1784 Sofia, Bulgaria \\
$^{4}$ Institute of Astronomy, Russian Academy of Sciences, 119017, Pyatnitskaya str., 48 , Moscow, Russia
}

\date{Accepted XXX. Received YYY; in original form ZZZ}

\pubyear{2023}

\begin{document}
\label{firstpage}
\pagerange{\pageref{firstpage}--\pageref{lastpage}}
\maketitle

\begin{abstract}
We present N-body simulations of the process of bulge formation in disc galaxies due to inward 
migration of massive stellar clumps. The process is accompanied by dark halo heating, with a quasi-isothermal core replacing the initial central density cusp, 
transforming an initially dark matter dominated central region into a baryon dominated one. The characteristics 
of the clumps are chosen to be compatible with low redshift observations of stellar 
clumps in DYNAMO-HST galaxies, which may be relatively long lived in terms of being 
robust against internal starburst-instigated disruption.  We thus test for disruption due to tidal stripping using 
different clump internal radial profiles; Plummer, Hernquist and Jaffe, in ascending order of steeper central 
density profile.  Our calculations predict that in order for clump migration to be effective in building 
galactic bulges and dark halo cores,  steeply increasing central clump profiles, or a less massive or less concentrated haloes, are preferred. The dependence on such factors may contribute to the diversity in observed total mass distributions and resulting rotation curves in galaxies. When the process is most efficient, a 'bulge-halo conspiracy', with a singular isothermal  total density akin to that observed bright galaxies, results. 

\end{abstract}




\section{Introduction}

As a model for cosmic structure formation, the standard cold dark matter (CDM) based scenario has developed into a particularly effective paradigm 
describing larger scales, but is still significantly lacking on smaller (sub)galactic ones 
(e.g. \citealp{FWRev2012, PrimackDD2012}).  
On  these scales there has  been 
no shortage in controversy and apparent challenges, 
including longstanding problems  relating to
the central densities of dark matter haloes and 
the properties of subhaloes in the local universe (see  \citealp{PrimackDD2012, DelpopRev2017Galax, BBKRev2017, SalucciRev2019} for reviews).  This consists of a triplet of possibly related issues.  

The first of these  concerns  the apparent discrepancy 
between the number of  
subhaloes predicted by CDM-only N-body simulations compared to  the number of satellite galaxies observed (the so called `missing satellite problem"; \citealp{Klypin1999,Moore1999}). The subhaloes  obtained in CDM-only cosmological simulations also appear too dense relative to those observed around the Milky Way (\citealp{Read2006MNRAS.367..387R, GarrisonKimmel2013,GarrisonKimmel2014}; but see however \citealp{2023arXiv230608674K}), with observed subhaloes lacking simulated counterparts. This has been labelled the
`too-big-to-fail" problem (\citealp{Toobigtofail2011}). 
The third issue is concerned with  
the discrepancy between the  centrally divergent dark halo density profiles obtained in  the simulations (\citep{nfw1996,nfw1997,Navarro2010})
with the  observations of dwarf spirals, dwarf spheroidals, and low surface brightness galaxies,  showing cored profiles (e.g., \citealp{Moore1994,Flores1994,Burkert1995,deBloketal2003,Swatersetal2003,DelPopolo2009,DelPopoloKroupa2009,2012MNRAS419971D,deBloketal2003,Swatersetal2003,DelPopolo2009, 2011ApJ...742...20W, 
DelPopoloHiotelis2014}). This is habitually referred to as the Cusp/Core problem, and constitutes a central aspect to be studied in the present work. 
It may also lie at the heart of all three aforementioned problems; as these issues may in fact be connected to the overdense self gravitating structures that the CDM 
scenario produces (\citealp{OgiyaBurkTBF2015}).

As the  small scale problems affecting CDM-based structure 
formation have been known for decades, several sets of solutions have been proposed, tackling the problems from different fronts. One set relates to the nature of 
the dark matter; popular alternatives to CDM, in this context, include warm dark matter (\citealp{2000ApJ542622C,2001ApJ551608S,BodeOst2001ApJ}), self interacting dark matter (\citealp{SpegelSteinSIDMPhRv,2000NewA5103G}), 
and fuzzy dark matter (\citealp{2000ApJ534L127P, BarkanaFDM2000PhRvL, Huietal2017}), as well as less studied scenarios invoking direct interaction
between the dark matter and baryonic components in the Universe 
(\citealp{FamaeyKhoury2020JCAP, SalucciInteract2020Univ}). 
Some improvements may also be obtained by modifying the  
power spectrum of primordial fluctuations
(\citealp{Kamionkowski:1999vp,2003ApJ59849Z}). 
Another possibility entails fundamental modification to gravitational theory (\citep{1970MNRAS1501B,1980PhLB9199S,1983ApJ270365M,1983ApJ270371M,Ferraro2012}), which naturally implies essential revision of the picture of the Universe 
envisioned through the currently standard $\Lambda$CDM model. 

As the problems arise at nonlinear scales, 
where complex baryonic physics becomes important, it was  
natural to consider the effect of the physics of the galaxy formation on the structure of the dark matter haloes. Solutions 
to the cusp/core problem, in that context, propose that the central densities of haloes 
may be reduced  as the baryons settle in the centres of 
haloes. 
For example, during a starburst supernovae driven gas can be blown out of the central regions, 
and the central potential may consequently decrease. CDM particles  
will then move to orbits of higher energies that take them farther away from the centre, giving rise to a decrease in the central density, and a flattening in the inner profile (as originally proposed by \citealp{Navarro1996a} and more recently studied 
by \citealp{Freundetal2020MNRAS, LiDeketal2023MNRAS}). 
Generally, a single blowout is insufficient to explain halo cores, and the gas may in any case re-enter 
the central region, but the process of dynamically 'heating' the central halo can be shown to be irreversible; leading to the production of a core due to repeated blowout (e.g., \citealp{DelPopolo2009, PonGov2012}; a wider description of the process can be found in section 2.2 of \citealp{DelPopolo2016}). As supernovae shock waves (and gravitational instabilities) may lead to fully developed turbulence in the feedback-driven gas, the process of energy transfer to CDM particles can also be envisaged as arising from potential fluctuations. 
connected to the density fluctuations of compressible turbulence. Repeated inflows and outflows can then be incorporated as part of a complete spectrum of perturbations
associated with stochastic fluctuations heating the dark halo (\citealp{EZFC,Haloheatin2023}). Full 
hydrodynamic simulations have indeed shown 
that potential fluctuations from feedback driven 
gas can indeed lead to core formation in the central region of CDM haloes, at least in the case of dwarf galaxies (\citealp{Gelato1999,Read2005,Governato2010,DiCintio2014, TeyssRead2013MNRAS}), but the efficacy of the process depends on the feedback 
recipe, and in particular requires a high star formation threshold
to maintain the sustained fluctuations leading to the dynamical 
heating of dark halo centres 
(e.g., \citealp{PGnat2014, 2019MNRAS.488.2387B, 
NIHAOThresh2019MNRAS, NIHAOthresh2020MNRAS}).

Another  proposed mechanism,  which originally 
inspired the repeated feedback-driven bursts models (\citealp{Mashchenko2006}), and is indeed closely related in terms of dynamics (\citealp{Haloheatin2023}), invokes dynamical friction-mediated  energy transfer to the dark halo particles from massive baryonic clumps during the galaxy formation process (\citealp{2001ApJ...560..636E, El-Zant2004}). 
In the standard CDM-based galaxy formation, gas condenses and cools inside 
dark matter haloes (\citealp{WhiteRees1978MNRAS}). If the gas distribution remains smooth during the process, adiabatic contraction ensues
(\citealp{Blumenthal1986,Gnedin2004}), further exacerbating the problem of central halo overdensity. 
This effect can nevertheless be overcome if the baryonic medium is clumpy
(\citealp{DelPopolo2009}), and the mechanism has been 
repeatedly shown to be effective, in principle, in reducing the central density of cuspy haloes into cores (e.g., \citealp{2008ApJ685L105R,Cole2011,2012MNRAS419971D,Saburova2014,Inoue2011,Nipoti2015}). 

The dynamical friction coupling mechanism may especially be called for in massive galaxies, 
as supernova feedback is generally insufficient for producing cores 
in larger galaxy haloes; indeed for  massive 
galaxies dark matter contraction due to baryonic settling wins out against supernova feedback driven core formation,  even in simulations that produce cores in small galaxies  (\citealp{Tollet2016MNRAS}. Even if AGN feedback can partially mitigate the effect;  \citealp{BardecontractMac2020MNRAS, BarDecontractFrenk2022MNRAS}). 
On the other hand, non-singular dark matter halo cores are  also inferred in massive low brightness galaxies (\citealp{deBloketal2001, deBloketal2003}) and may possibly also exist in massive luminous ones (\citealp{SalucciRev2019}). The dynamics of such galaxies seem especially inconsistent with observations if 
the sole effects of baryons on an NFW halo is modelled is conceived in terms of adiabatic contraction (\citealp{McGauadiab2022}). 
Although, in general, both feedback and dynamical 
friction induced halo heating may act in concert during galaxy formation \citep{orkneyJ2021}, dynamical friction mediated halo heating may be especially relevant in way of explaining these results, as well as those of apparently cored haloes of higher redshift massive galaxies (\citealp{dekelF2021,ognag2022}). 

When that mechanism was first proposed 
there were no observations of baryonic clumps of the required mass. For example, for a  Milky 
Way like galaxy, the required clumps were predicted to have masses $\gtrsim 10^8 M_\odot$, while the largest molecular clouds observed in local galaxies were an order of magnitude 
smaller in mass (and also not believed to be sufficiently long lived). 
There were observations of irregularly shaped and clustered early galaxies, but these were primarily interpreted in terms of merging and interacting systems of similar scale rather than  collections of compact clumps embedded in  larger dark matter hosts (\citealp{vdBergh1996}; even if the latter situation may also, in principle,  materialise in the context of hierarchical structure formation).  
However, since then, an increasing number of galaxies 
have been discovered to host massive and compact clumpy structures of the 
right mass. These clumps, which are more abundant at higher redshifts,  
may collapse {\it in situ} within a forming galaxy (\citealp{Elmegreen2004,Elmegreen2007, 
Elmegreen2009,Genzel2011,Schreiber2011,Guo2015,Guo2018,messaCombes2022,martinA2023, 
Sattari2023}).  
Their existence may be related to disc formation and can potentially have important consequences regarding the the structure of the host galaxy; 
when gas is infalling towards a forming disc,  radiative cooling induces  self-gravitating instabilities that lead to the formation of massive clumps, which may in turn migrate --- due to dynamical friction, as well as to torques arising from scale perturbations in the disc --- to the centres of their host galaxies (\citealp{Noguchi1999, Agertz2009, Ceverino2010, mandelker2014, mandelker2017, Dekel2022}), possibly removing dark matter from the vicinity of a nascent bulge in the process
(\citealp{ElmB2008}). 
It has also been suggested that instabilities in purely stellar discs may lead to clumpy structures that are sufficiently long lived to likewise form a central bulge
(\citealp{SahaSanakStellarClumps2018ApJ}). 
Though there are varying views on the matter (cf. Section 2.2.3 of \citealp{DelPopolo2016} for a discussion), some observational studies are consistent  with the expectations of an evolutionary  scenario invoking sufficiently long lived (> 100 Myr) clumps forming {\it in situ} and migrating to the centres of galaxies 
(\citealp{Genzel2008,Genzel2011,Shibuya2016,Soto2017,Zanella2019}).

Nevertheless, the lifetime of  of high redshift clumps, particularly primarily gaseous ones, is still uncertain; they may indeed disrupt over 
timespans much shorter than the of inward migration 
timescale (\citealp{murray2010, oklopfire2017}). 
The fragmentation mass scale found in simulations
may  furthermore be subject to numerical resolution effects (\citealp{Turbores2015}). Resolution may also affect 
the observationally inferred effective radii and stellar masses of the clumps (e.g., \citealp{miroslava2017res1,FisherDyn2017, miroslava2018res2, Faure_Bournaud2021}). 
Selection effects related to the limited rest frame wavelength range available to the HST may further affect the inferred masses.  Thus, although early studies using the JWST seem consistent with clumps surviving for up to a Gyr and centrally migrating  (\citealp{JWST2023pap1, JWST2023pap2}), it is still not obvious that observed higher 
redshift clumps  have the right 
properties (mass, size and robustness) 
for dynamical friction coupling with the host system to be sufficient for bulge building or halo core formation.   

On the other hand, local analogues of early clumpy galaxies, albeit rare, do exist and 
have been studied (e.g. \citealp{FisherDyn2017, Fisher2017, messa2019loc}). 
The present work is an attempt to make use of such observations to pin down and isolate the potential importance and consequences of dynamical friction mediated coupling in clumpy galaxies, 
To this end, we undertake N-body simulations of stellar clumps embedded in a disc-halo system, with clump masses and sizes consistent with those presented the relatively well resolved DYNAMO-HST sample (\citealp{2022MNRAS.512.3079A}).
If sufficiently long lived, 
such observed clumps would be in the process of migrating 
towards the centres of their host galaxies, a process that may lead to the 
simultaneous formation of a central bulge (from the remnants of the 
clumps), in addition to a halo core replacing the cusp. Thus, even galaxies that are initially 
dark matter dominated at all radii may end up with much less dark matter in their central regions; as the dark matter
is replaced  by a nascent baryonic bulge.  
Such a scenario 
appears particularly 
appealing as it may contribute to the 'conspiring' 
of dark halo and baryonic components to maintain flat 
rotation curves, a phenomenon sometimes termed as the disk halo or bulge halo conspiracy. And, as our results here also suggest, to the wide
diversity of observed galaxy rotation curves.

Whether this scenario  actually materialises in galaxies  will 
depend on the timescale associated with the clump migration relative to the lifetime of the clumps. 
The latter timescale 
in turn entails two processes: 
disruption of the clumps as a result of further star formation, and tidal stripping due to the motion of clumps 
in the disk-halo potential of the host galaxy. 
The internal composition and density profile of the DYNAMO-HST clumps, for which only stellar masses 
are available, is still uncertain. 
However, the work of \cite{2021MNRAS.506.3916L} suggests that the clumps have a complex substructure of stellar populations with different ages, with the centres of the clumps being populated with young stars while older ones occupy the outer parts. Accordingly, and given their age distribution as a function of distance from their galactic centres, the DYNAMO-HST clumps were interpreted as relatively long-lived structures; in the sense of being robust against sudden, internal starburst-induced disruption.   
We thus focus here on the second effect, namely that of disruption due to tidal stripping. 

We thus conduct numerical simulations of disc galaxies containing stellar clumps of similar masses and sizes, and radial distribution in the host system,  to those found in the  DYNAMO systems. As the internal density profiles of the DYNAMO clumps 
is uncertain, we try three different theoretical ones
(Plummer, Hernquist and Jaffe profiles), to test under what 
conditions the clumps may survive long enough to centrally migrate to form a stellar bulge and induce a dark matter core. 
We also examine the variety of rotation curve shapes that 
arise from the process, and how it depends on host halo mass and concentration.  
These results are described in Section~\ref{sec:results}. In the next section we first outline the properties of the simulated clumps, and how they are related to the observations, while in Section~\ref{sec:setup} we describe the simulation setup, including the effective initial conditions for our disc-halo-clump system. We summarise and discuss our results in Section~\ref{sec:conc}.

\section{The DYNAMO-HST stellar clumps}
\label{sec:clumpsobs}

To obtain observationally constrained models of stellar clumps embedded in galactic discs, we use the recent DYNAMO-HST observations of the stellar masses of baryonic clumps, as presented in~\cite{2022MNRAS.512.3079A}. 
These involve  six galaxies from the DYNAMO sample (\citealp{2014MNRAS.437.1070G}), which  are selected from Sloan Digital Sky Survey (SDSS) DR4~(\citealp{2006ApJS..162...38A}). The galaxies are local ones, with redshift range $z=0.075-0.13$, exhibiting a turbulent 
gas rich bulge-less disk with high star formation rate. 
The inferred stellar masses of the galaxies  range from  $1.7\; {\rm to }\; 6.4 \times 10^{10} M_{\odot}$, the upper limit being close to that of the Milky Way
(\citealp{MassMY2015ApJ}).

\subsection{Clump sizes and masses}

In~\cite{2022MNRAS.512.3079A}, the size of the identified clumps are measured by fitting an elliptical Gaussian function to the 2D brightness distributions of the clumps around the peak. The effective radius of each clump is then defined as the mean standard deviation of the major and minor axis of the best fit 2-D Gaussian functions.  This is the region inside which a young stellar population of the clumps is exceedingly bright compared to the background discs they are embedded in.  The stellar masses of the clumps are then inferred by fitting stellar population synthesis models to the observed spectral energy distributions, distinguishing between 'raw' and 'disc subtracted' clump masses. 

\begin{figure}
	\includegraphics[width=\columnwidth]{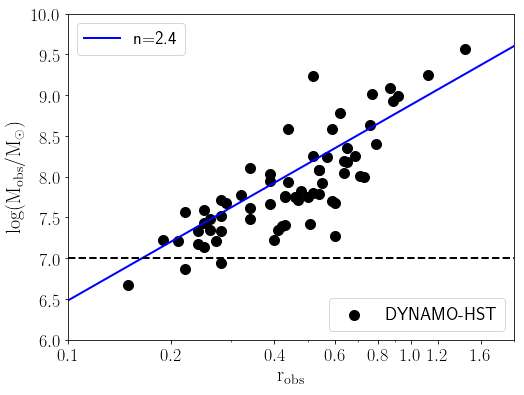}
   \caption{The DYNAMO-HST clumps mass-size relation (from the data of ~\citealp{2022MNRAS.512.3079A}).  
   The solid line is the power law fit to the mass-size scaling. The horizontal black-dashed line represents the observational mass limit of $10^7  M_\odot$.
   The masses are inferred by fitting the radial light distribution by 2-D Gaussians, with cuts on orthogonal axes corresponding to one standard deviation (which corresponds to a fraction of about 0.39 of the region under such a curve, assuming equal standard deviations on both axes). To construct physical self gravitating mass models,   
    we associate the effective radius, thus inferred, with the half mass radius of clumps with total masses $M_c = 2 M_{\rm obs} = 2 M_{1/2}$, and with internal mass distributions modelled through three different physical density profiles (Section~\ref{sec:clumpsprof}).
   }
    \label{fig:masssize} 
\end{figure}

As the clump sizes are of order of the disc thickness,
and as we will assume the clumps to be self-gravitating and dynamically distinct objects, we consider only the raw masses, which 
we refer to here as $M_{\rm obs}$. 
These   range from $4.71\times 10^6$ to $3.74\times10^9 M_{\odot}$, with average mass $2.53\times 10^{8} M_{\odot}$. We note that they are about an order of magnitude smaller 
than stellar clump masses quoted in many higher redshift studies, an effect that 
may arise from the lower resolution available in these studies (\citealp{FisherDyn2017})
The observed clump mass-size relation may be fitted with a power law, $\rm M_{\rm obs} = r^n_{\rm obs}$, with $n = 2.4$, as illustrated in  Fig.~\ref{fig:masssize}.

As mentioned, the masses of the clumps are observationally evaluated by 
fitting their light curves with Gaussians and imposing 
a cutoff at one standard deviation.
We will want to construct physical self gravitating mass models for the observed clumps. 
As a 2-D
standard deviation (with same deviation in each dimension) corresponds to a fraction of 0.39 times the light comprised under the full 2-D Gaussian curve, we will generally assume that this cutoff --- and the associated observationally quoted effective radius for each clump --- corresponds to the half mass and half mass radius of the clumps. For which we will 
adopt various mass models to describe (Section~\ref{sec:clumpsprof}).

\subsection{Clumps mass function}

According to the data given in~~\cite{2022MNRAS.512.3079A},  the average number of clumps per galaxy in the DYNAMO sample 
is about 13, which is the number we adopt in each of the simulations 
described below. The 
number of clumps as a function of  mass may be fit by the following mass function 
\citep{2022MNRAS.512.3079A}:
\begin{equation}
    \phi(x) = \ln(10) \times \left(\frac{M_{\rm obs}}{M_0}\right)^{\alpha + 1.0},
    \label{eq:massfun}
\end{equation}
where $x = \log(M_{\rm obs})$, $\alpha = -1.4$ and $M_0=10^{9.5} M_{\odot}$.
As a realisation of this mass faction, we take four equal logarithmic mass bins in which the clumps are distributed. 
In accordance to the discussion above, the 
total mass of each clump $M_c$
is assumed to correspond to twice 
the masses $M_{\rm obs}$, which are taken 
to correspond to the half mass of the clumps.  
In practice, in order to arrive to a quasi-equilibrium of the simulated halo-disc plus clump system, while approximately matching the masses $M_c$ and radial location $R$ of the clumps in the observed sample, we start the our simulated clumps 
at large $R$  (relative to the observed ones), with masses $2 M_c$, and let the system relax, with clumps migrating inwards through dynamical friction and losing mass due to stripping, for about a rotation period. This procedure, which sets our effective initial conditions, is described in Section~\ref{sec:ClumICs}.

\section{Simulation Setup}
\label{sec:setup}

We advance the dynamics of the disc, halo and clump system, with properties and parameters described below,  
with the publicly available GADGET-4 \citep{2021MNRAS.506.2871S} code for N-body simulation of isolated systems. The code employs fully adaptive 
time-stepping and uses a hierarchical tree algorithm, in combination with a particle-mesh  scheme, 
to  adaptively compute the gravitational forces. 

The simulations involve two types of particles; dark matter and stellar, composing the halo and the disc with its embedded clumps, respectively. The gravitational softening of both species is  set equal to $\epsilon = 0.035\; \rm kpc$. The maximum simulation time is set to $\rm 3\; Gyr$. In the runs 
described below, the total number of particles
composing the halo  
is 
$N = 256^3$, with mass resolution of $m_p = 5.9\times 10^{4} M_\odot$. 
Convergence and initial equilibrium tests 
are presented in Appendix ~\ref{app:conv}

In the next subsections we describe the structure of the components of our disc-halo-clump 
system. Then we detail our starting initial conditions, and the  relaxation towards a quasi-equilibrium, with clump positions and mass 
distribution consistent with the DYNAMO observations, which will represent 
the effective initial conditions for our runs. 

\subsection{The halo and the disc}
\begin{figure}
    \centering
    \includegraphics[width=\linewidth]{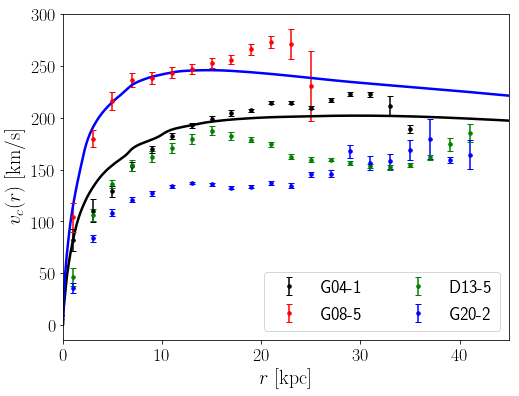}
    \caption{Available Observed rotation curves of DYNAMO-HST 
    galaxies used in this work (\citealp{2019MNRAS.485.5700S}),
    versus simulated ones, taken at the effective
    start of the simulations (at $T=0$, after a 500 Myr relaxation period as described in Section~\ref{sec:ClumICs}). The theoretical lines show models with halo concentrations $c=20.4$ and $c=10.2$.}
    \label{fig:DYNAMORCs}
\end{figure}

The N-Body simulations presented  here are of clumps embedded in an isolated galaxy composed of: 
\begin{enumerate}
    \item 
    A dark matter halo following a Navarro,Frenk and White (NFW) density profile \citep{nfw1997}, 
    \begin{equation}
        \rho = \frac{\rho_s}{\frac{r}{r_s}\left(1 + \frac{r}{r_s}\right)^2},
        \end{equation}
where $\rho_s$ and $r_s$ are density and radial scales.  

In our fiducial model the total host (disc + halo) mass is 
$10^{12} M_\odot$, so that the halo mass 
is $(10^{12} - f_d) M_\odot$, where $f_d$ is the disc
 mass fraction (taken as 0.02 in our fiducial model). We assume a  flat-$\rm \Lambda CDM$ cosmology (with Hubble constant $H_0 = 70.0$ and matter density ($\Omega_m = 0.3$), so that  the halo viral radius is  $r_{\rm vir} \equiv c r_s = 204.0~{\rm kpc}$, where $c = 20.4$ is the halo concentration. This  is larger than the average found in simulations at this 
mass scale (which is closer to $c=10$ at low redshift; e.g., \citealp{Pradetal2012}),
but it ensures that the model galaxy is initially entirely dark matter dominated at all radii, including the central region.
One of our goals will be to 
probe the transformation from such 
a situation into one whereby the 
centre becomes baryon dominated, which occurs concurrently with the formation of the halo core.  
We also run models with halo concentration parameters 
$c=10.2$ for comparison.
In all cases, to avoid a sharp cutoff, our halos are simulated up to radius larger than the assumed virial radius  (namely up to $r = 230~{\rm kpc}$). In the absence of the clumps,  
halo profiles have been checked to, remain unchanged (after reaching equilibrium state) during the simulation time, as shown in Fig. \ref{fig:HostDensProf}.

\item A stellar disc, with density profile given by:
\begin{equation}
    \rho_d (r) = \rho_d \exp\left(\frac{-l}{l_d}\right)  \sech^2\left( \frac{z}{z_d} \right), 
\end{equation}
where $l = \sqrt{x^2 + y^2}$, $l_d$ is the radial disc density scale 
and $z_d$ is the vertical scale height. 
To ensure stability against significant non-axisymmetric instabilities, such as strong bars, we choose a large value for the Toomre parameter, namely $Q = 2.0$.
\end{enumerate}

The values of the profile parameters of each component of the host system in our fiducial system are listed in Table \ref{table:hostprop}.  For the radial extent of interest to us (less than 10 kpc), our starting mass distributions, for $c = 20.4$ and $10.2$, approximately 
correspond to the range spanned by the 
three more massive galaxies of the four galaxies  
in the DYNAMO samples used used here for which 
rotation curves are available, 
as may be seen from from Fig.~\ref{fig:DYNAMORCs}. 
We also ran simulations  with less massive halos and more massive discs (such as the one corresponding to the rotation curves in Fig.~\ref{fig:conspiracy}).

\begin{table}
    \centering
    \begin{tabular}{l|c|c}
         & Halo & Disc \\
        \hline
         $M~[M_{\odot}]$ & $(1-f_d) \times 10^{12}$ & $f_{*} \times 10^{12}$ \\
         $r_s~~{\rm [kpc]}$ & 10.0 & -- \\
         $c$ & 20.4 & --\\
         $l_d~~{\rm [kpc]}$ & -- & 3.5 \\
         $z_d~{\rm [kpc]}$ & -- & 0.5 \\
         \hline
    \end{tabular}
    \caption{The fiducial host (disc + halo)  mass distribution parameters. The total mass of the disc and halo combined  is $\rm M_{\rm host}  = 10^{12} M_{\odot}$, and the disc mass fraction is $f_d = 0.02$. The exponential disc has vertical and radial scale length $l_z$ and $l_d$, and the NFW halo has concentration $c$ and radial scale length $r_s$. 
    In addition to those fiducial parameters we ran simulations with less concentrated halo, with $c=10.2$, and also ones where the halo mass is halved and disc mass doubled.}
    \label{table:hostprop}
\end{table}

\subsection{The internal profiles of simulated clumps}
\label{sec:clumpsprof}
We add to the halo-disc system stellar clumps with properties  constrained 
by the DYNAMO-HST observations of \cite{2022MNRAS.512.3079A} (as described in Section\ref{sec:clumpsobs}). 
The resolution of the observations did not permit  
detailed inference of the clump density profiles. 
Indeed, as mentioned, the observed radial light profiles of the clumps were simply fit by a Gaussian and cut at one-sigma, 
which we take to constrain the half mass.  

A primary 
goal of the current study is to determine 
the robustness of the stellar clumps against striping, 
which may determine the possibility of the inward migration of the clumps, 
the building of bulges from their remnants, 
and the formation of halo cores from dynamical friction mediated 
energy transfer to dark matter particles. 
As these processes are expected to strongly depend on 
the internal structure of the clumps, we try three different profiles, differing in the steepness of their radial central density distribution. 

The three spherical density distributions we consider consist of: i) 
a Plummer profile, characterised by a central 
nearly constant density core; ii) a Hernquist profile, with central density rising as $1/r$; iii) a still more cuspy Jaffe profile, with density going as $1/r^2$ in the central 
region. The associated mass distributions are  
    \begin{equation}
    M(< r) = 4 \pi \rho_{c} r_{c}^3 \begin{cases}
    
    \frac{\left(\frac{r}{r_{c}}\right)^2}{2\left(1 + \frac{r}{r_{c}}\right)^2}  ,  & \text{Herquist} \\
    
    \frac{\frac{r}{r_{c}}}{1 + \frac{r}{r_{c}}}, & \text{Jaffe}\\
    \frac{1}{3}\frac{r^3}{2^{-\frac{3}{2}}\left(r^2 + r^2_{c}\right)^{\frac{3}{2}}}, & \text{Plummer}
    \end{cases}
    \label{eq:NFWProf}
\end{equation}
where $\rho_{c}, r_{c}$ are density and radial scales characteristics of the clumps. These are fixed through the 
observationally inferred 
half mass radius and the  
tidal truncation radius described below. 
The  (isotropic) internal velocities  are set by the spherical Jeans equation.

\subsection{Tidal truncation of the clumps}
\begin{figure}
\begin{center}
\includegraphics[width=  7 cm]{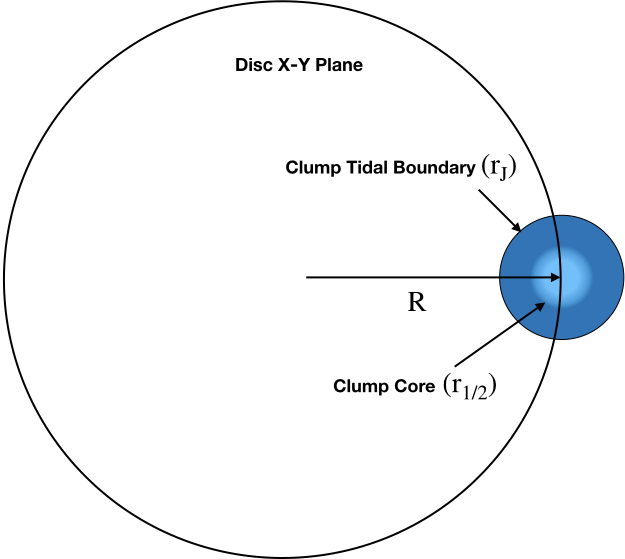}
 \end{center}
    \caption{Internal clump profile sketch. In order to determine the scale parameters of the profile, we set constraints on the clump core mass and effective radius from DYNAMO-HST observations, and  define the  clump boundary  through the tidal (Jacobi) radius at the clump's orbital position in the host's disc plane.}
    \label{fig:ClumProfSketch}
\end{figure}

\begin{figure}
         \includegraphics[width= 8 cm ]{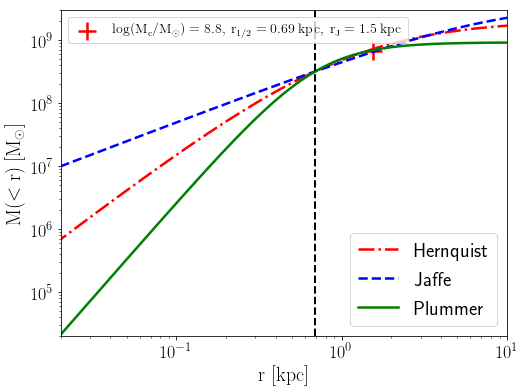}
    \caption{The different mass profiles of a clump with mass $M_c = 10^{8.8} M_\odot$. The vertical black-dashed line is at the half mass radius, which we take to be fixed by observations. This, along with the Jacobi tidal truncation radius completely fixes the internal clump density distribution, given the profile.}
    \label{fig:DiffMassProf}
\end{figure}

    None of the profiles just described have finite 
    truncation radii, as the associated densities are finite 
    at any finite radius from the centre. Clumps embedded in our disc-halo system, however, will be tidally truncated.  We use the Jacobi radius to approximate the expected  tidal truncation scale. For a clump of mass $M_c$ and orbital radius $R$, this is defined as \citep{2008gady.book.....B}:
    \begin{equation}
    \rm r_{J} = \left(\frac{M_c}{3M_{\rm host}(<R)}\right)^{1/3}  R,
    \label{eq:jacobiRd}
\end{equation}
where 
$M_{\rm host}(<R)$ is the enclosed mass of the host (that is disk plus halo mass) at radius $R$
in the disc plane.
The clump profile parameters are then determined from the 
condition that the masses of the clumps be completely contained 
within $r_J$ and that half the mass be contained within
the observed radius $r_{\rm obs} = r_{1/2}$.

\begin{figure*}
\includegraphics[width= 8.5 cm ]{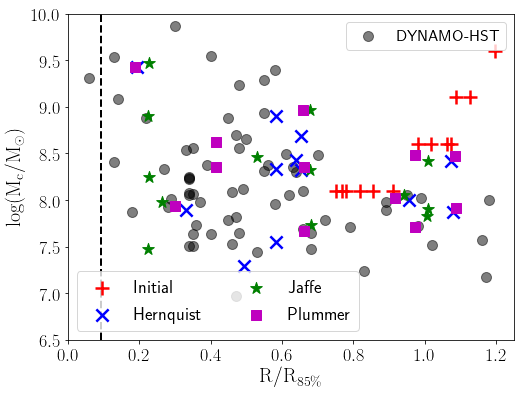}
	\includegraphics[width= 8.5 cm]
 {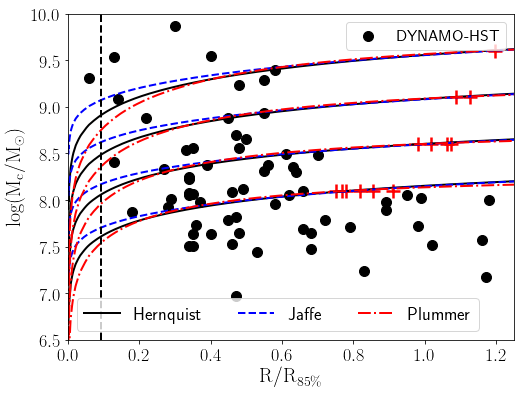}
    \caption{Setting the effective initial conditions. The initial (at $t=0$) clump masses are distributed in logarithmic mass bins in accordance with equation (\ref{eq:massfun}), as described in Section~\ref{sec:clumpsobs}.  The masses  are then doubled and the clumps are placed at large radial positions $R$ in the disc (relative to those observed), corresponding to the red crosses on the right of the plots. The left panel shows  the positions of the clumps after the (disc-halo-clumps) system is left to relax by advancing the combined dynamics for a relaxation period of 500 Myr (corresponding to $T=0$), with different 
    symbols denoting runs with the three different internal clump profiles used. The grey round dots correspond to the actual masses and positions of the clumps (in all combined) DYNAMO galaxies. On the right hand panel, the lines represent the estimated remnant mass, when the tidal stripping is estimated using eq. \ref{eq:jacobiRd}. The vertical black dashed line is at $\rm R = 1 {\rm kpc}$, where the remnants of larger clumps are expected to merge into a bulge after completing their central migration.  The scaling radius $R_{85 \%}$ refers to the observed disc radius comprising $85 \%$ of the light, which we assume to correspond to the same percentage in mass of the simulated discs.}
    \label{fig:ClmpInitDist}
\end{figure*}

Fig.~\ref{fig:ClumProfSketch} shows a sketch illustrating the two radii 
that thus fix the internal clump mass distribution, namely the half mass radius $r_{1/2}$ and the truncation radius $r_J$.
As a quantitative example of the resulting mass profiles, 
Fig.~\ref{fig:DiffMassProf} shows those corresponding to  
a clump of half size 
$r_{1/2} = 0.27\; {\rm kpc}$  and total mass $M_ c = 10^{8.8} M_\odot$, located at an orbital radius $R  = 9.64~{\rm kpc}$ in the disc plane.

\subsection{Quasi-equilibrium start and relaxation to effective initial conditions}

\subsubsection{Disc halo system}

The quasi-equilibrium starting conditions for the disc-halo system are created using the Disk Initial Conditions Environment (DICE) code \citep{2014A&A...562A...1P, 2016ascl.soft07002P}; the particle distribution is thus created using an N-try MCMC algorithm, which does not require prior knowledge of the distribution function. The dynamical equilibrium of each component in the system is computed by integrating the Jeans equations for each disc and halo particle. 

\subsubsection{Clumps}
\label{sec:ClumICs}

\begin{figure}
	\includegraphics[width=\linewidth]{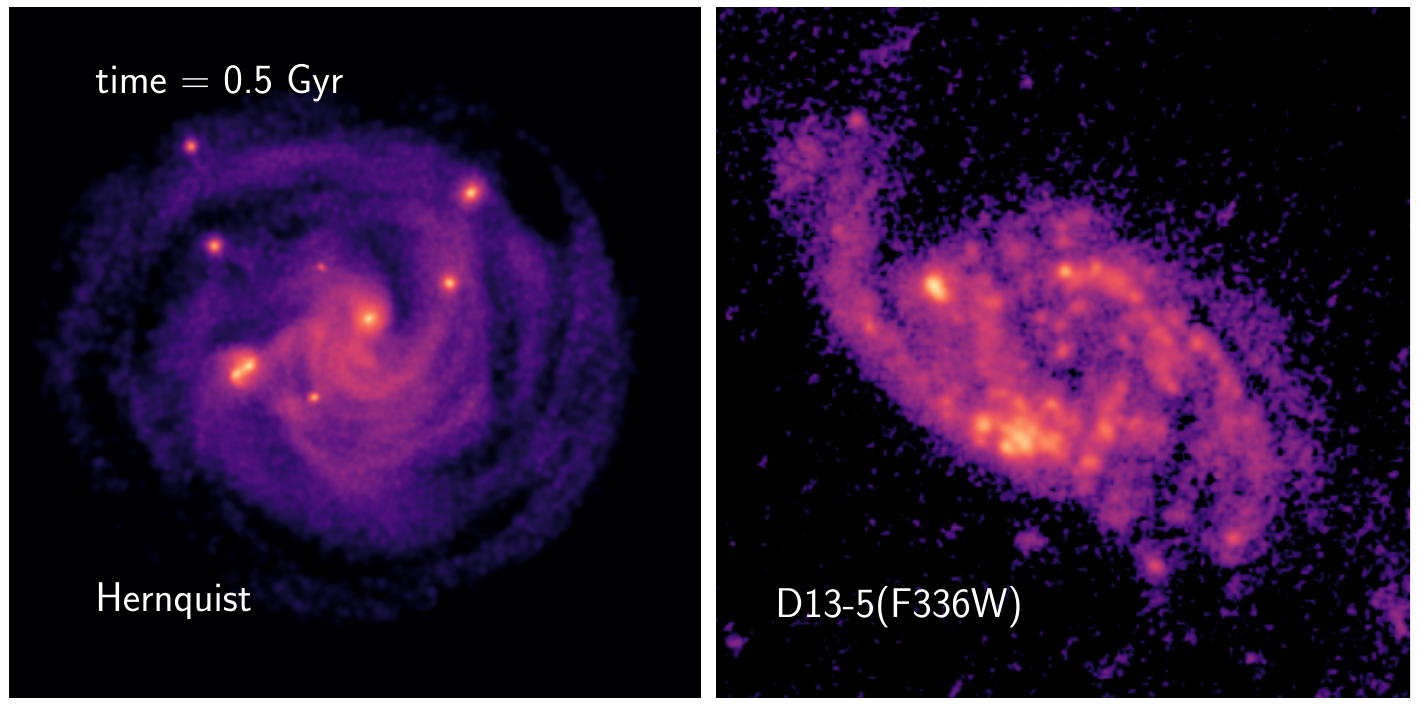}
    \caption{Left: Simulated  
    galaxy with clumps of Hernquist internal mass profile after the initial  relaxation period comprising $500 {\rm  Myr}$ of evolution ($T=0$). Right: DYNAMO-HST galaxy D13-5 using F336W filter.}
    \label{fig:DiscClmpInitDist}
\end{figure}

\begin{figure*}
        \includegraphics[width=\linewidth]{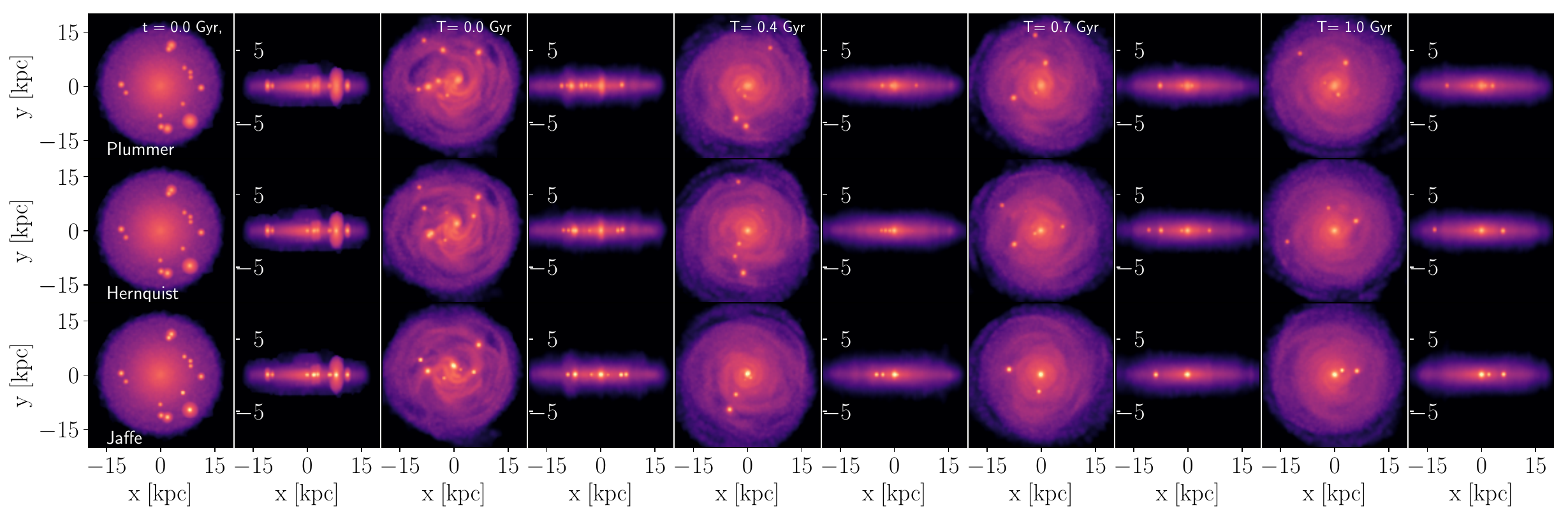}
    \caption{Projected density maps of the stellar disc and clumps, for simulations of the fiducial model (Table~\ref{table:hostprop}), with different clump profiles, as noted. The clumps move to the centre of the disc-halo host due to dynamical friction coupling. The resulting bulge size differs depending on the clumps' internal profile.  
    The time $t = 0.5~{\rm Gyr}$, corresponding to $T = 0$, denotes the effective starting conditions of our simulations, when the models relax to a quasi-steady state with clump positions and masses in approximate correspondence with the observed ones (cf. Fig~\ref{fig:ClmpInitDist}).}
    \label{fig:DiskClumpImg}
\end{figure*}

Each clump centre of mass is assumed to initially move 
at the local circular speed at its starting position in the disc-halo system. The combined system is only in initial approximate quasi-equilibrium 
and must therefore be advanced for a time of the order of a 
rotation period in order to settle. 
This is also necessary for obtaining 
stable clump masses, as the initial internal 
dynamical equilibrium of the embedded clumps, as well as their initial (tidally truncated) radius can only be approximate, since the Jacobi relation (eq. \eqref{eq:jacobiRd}) for the the tidal radius is approximate. Indeed, as shown in Appendix~\ref{app:Jacoby}, it is also a 
better approximation when the clumps are at larger $R$ than if used directly to truncate the clumps at smaller radii within the disc.

We adapt the following procedure so as to arrive at a mass and positional 
distribution of clumps compatible with that observed in the DYNAMO galaxies 
after an initial 'relaxation' period. 
We initially set the positions of the clumps on the plane of the disc at relatively large 
radii, namely with $8.0 < R < 16.0 {\rm kpc}$, with a normal distribution for their radii and random azimuthal angles, and imposing the condition that the Jacobi radius $r_J$ of each clump is bigger than its half mass radius $r_{\rm obs}$
(which naturally orders them in radii $R$ according to their masses). 

As the clumps migrate inwards from their initially large radius (due to dynamical friction), they also lose 
mass due to stripping. 
To match the observed mass-position relation \citep{2022MNRAS.512.3079A} of the clumps, as depicted in Fig. \ref{fig:ClmpInitDist}, we  double the initial mass and evolve 
the system for 500 Myr. This ensures that the observed 
distribution is approximately matched, and also that the system reaches a quasi-steady state, 
which may be assumed to correspond to the effective
start of the simulation, We denote this time with $T = 0$.

We illustrate the procedure in Fig.~\ref{fig:ClmpInitDist}.~~\footnote{If anything, the applied procedure somewhat underestimates 
the preponderance of massive clumps in the central region, 
and in turn the associated effects due to dynamical friction 
coupling that we describe below.}
Fig.~\ref{fig:DiscClmpInitDist} exhibits a pictorial comparison 
of the effective initial conditions in the simulation with one 
of the  observed DYNAMO-HST galaxies. 
Finally, it is important to note that, as we will see below, the physical changes in the system 
we are looking for, namely halo core and baryonic bulge formation, are absent during the relaxation period. 
This may be expected, as the relaxation process process 
involves the migration of clumps through the outer region of the system. The effects we are interested in, on the other hand,  occur with the subsequent migration to the inner parts, which is described below.

\section{Results}
\label{sec:results}

Dynamical friction coupling between the  massive stellar 
clumps and the host system leads to clump migration 
toward the centre. This is illustrated in Fig.~\ref{fig:DiskClumpImg}, where 
we show  projections of the evolution of the disc-clump 
system. As the clumps migrate towards the centre they lose energy to the 
background, composed of the disc and the halo, which is thus 
dynamically 'heated'. As the sub-system of halo particles is more centrally concentrated, it has larger binding energy and therefore 
absorbs most of the energy lost by the clumps.
This results in a central halo core replacing the initial density 
cusp. The remnants of the inwardly migrating clumps 
accumulate to form a central bulge. 
The end result of these processes is the once dominant dark matter becoming  insignificant 
in terms of contribution to the central rotation 
curve, with the main contribution there coming from the baryons by the end of the evolution. We describe these processes in more detail below.

\subsection{Energy transfer from clumps to cusp and the transformation to quasi-isothermal core}

\begin{figure}
	\includegraphics[width=\linewidth]{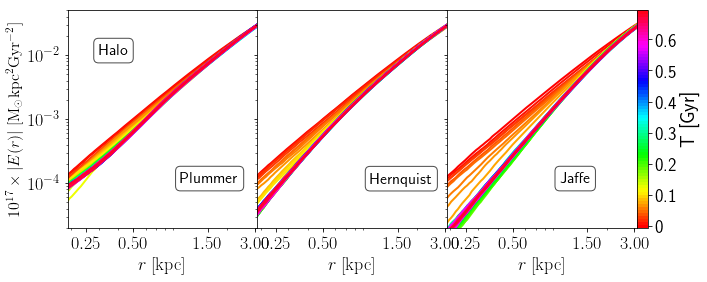}
    \caption{Change in total energy of subsystems of halo particles enclosed within radius $r$, as given 
    by equation~\ref{eq:Eind}, as they gain energy from the inwardly migrating stellar clumps. The results are illustrated for the fiducial disc-halo model of Table~\ref{table:hostprop}, and for the three internal clump profiles considered: Plummer spheres with constant density cores; Hernquist spheres, with $1/r$ inner cusps; and Jaffe  spheres, 
    with $1/r^2$ inner cusps.}
\label{fig:haloen}
\end{figure}

The energy gained by the inner halo as the clumps
migrate inwards is illustrated in Fig.~\ref{fig:haloen}, which shows the energy of systems made of subsets of halo particles enclosed within spherical radii $r$. This is defined as  
\begin{equation}
 E_r = \frac{m_p}{2}  \left( \sum_i v_i^2  +  \Phi_i \right),
    \label{eq:Eind}
\end{equation} 
where $m_p$ refers to the mass of a  particle (recall that all particles 
have the same mass in our simulations),  
$v_i$ is the speed of particle $i$ and $\Phi_i$ is the Newtonian potential (due to all other particles 
in the simulation), 
at its location. The summation 
is evaluated over all halo particles within radius $r$.

One immediately sees that the energy input rate is larger for the more concentrated internal clump mass distribution; namely for the 
Jaffe profile, with its steeply increasing central density cusp.  
This is due to the fact that as the clumps spiral to the centre they lose mass 
due to stripping, a process that is least efficient  
for steeper internal clump profiles. As the energy loss 
rate is proportional to the square of the mass, this is an 
important effect. 
The competition between energy transfer and stripping in fact determines the efficiency of the
energy transfer from the stellar clumps to the background particles, and associated the erasure of the halo cusp as we will see below.

\begin{figure}
	\includegraphics[width=\linewidth]{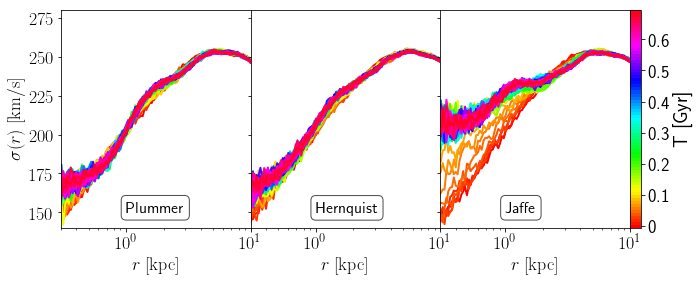}
    \caption{Evolution of the (three dimensional) velocity dispersion profile of the host halo, for the same runs as in Fig.~\ref{fig:haloen}.}
\label{fig:halovely}
\end{figure}

The halo central density cusp is 
particularly sensitive to the energy input because it suffers from 'temperature inversion', with inner velocity dispersion decreasing 
towards the centre, which renders its structure  
susceptible to even small 
energy input; as even modest gain in kinetic energy is significant given 
its initially small value. Thus, as the halo particles gain energy from the 
in-spiraling clumps, an isothermal core starts to  
replace the temperature inversion connected to the initial cusp
(Fig.~\ref{fig:halovely}). Again,  the effect is most prominent in the case with the most centrally 
concentrated clumps (with internal Jaffe profiles).

\begin{figure}
\includegraphics[width= 8.5 cm ]{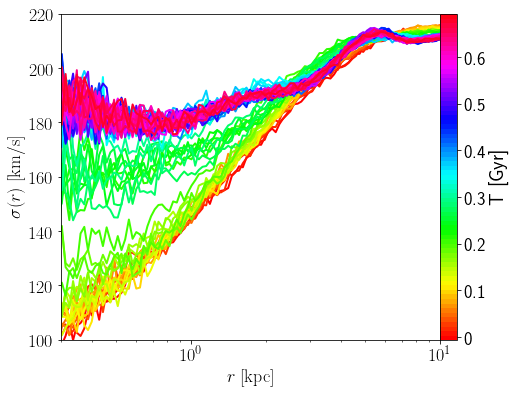}
   \caption{Time evolution of Velocity dispersion  when the halo concentration is $c=10.2$, instead of $20.4$. The internal clump profile is of the Jaffe form.}
   \label{fig:halovely_jaffC10}
\end{figure}

When a less concentrated halo profile is used 
(initial NFW $c=10.2$ instead of 20.4), 
halo particles are less bound, and the central halo structure is thus more susceptible to energy input.  
This results in a larger effect in terms of total energy change, as well as increase in velocity dispersion that leads to the tendency towards an isothermal core (Fig.~\ref{fig:halovely_jaffC10}).

\subsection{Density profiles}

\subsubsection{Cusp-core transformation}

\begin{figure}
	\includegraphics[width=\linewidth]{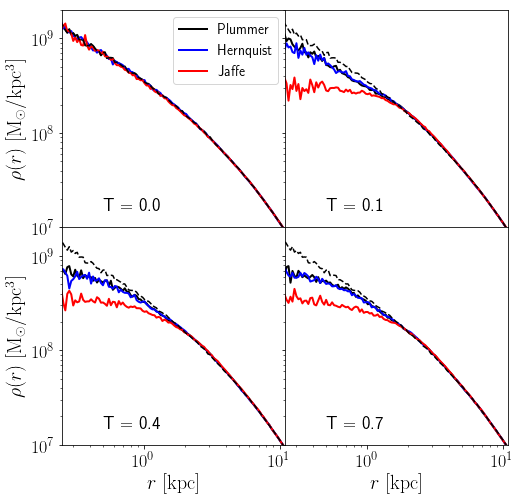}
    \caption{Halo density profile for the fiducial runs (corresponding to those in  Fig.~\ref{fig:haloen}).} 
    \label{fig:HaloDensProf}
\end{figure}

The energy transfer  through dynamical friction, and the 
accompanying advent of a quasi-isothermal inner velocity dispersion profile, 
leads to the replacement the initial 
halo density cusp by a core. 
In Fig.~\ref{fig:HaloDensProf}, we plot the halo density profile in our fiducial model for the different 
internal clump profiles. 
After a few hundred Myr of evolution from our initial relaxed quasi-steady state,  a nearly constant density core, of order of a kpc in scale, replaces the central cusp. Given the considerations regarding the efficiency of the stripping as a function 
of clump internal profile, the precise timescale and the core size depends on that profile.  The process being again more effective for more steeply increasing clump central density.

\begin{figure*}
	\includegraphics[width= 8.5 cm ]{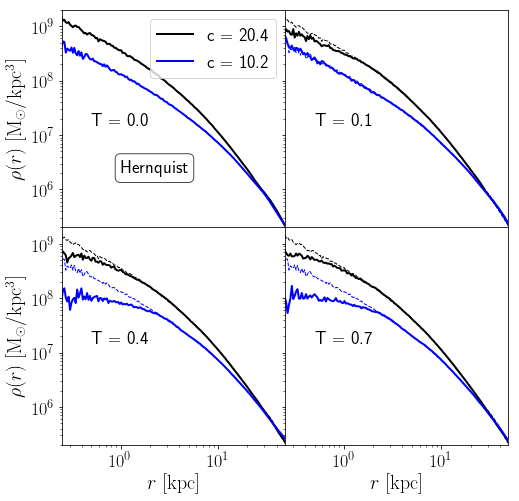}
 \includegraphics[width= 8.5  cm ]{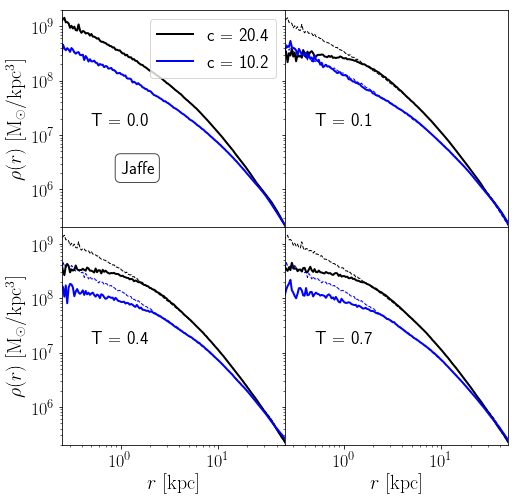}
   \caption{Comparing the evolution of host halo density profiles when the initial  NFW concentration is $c=10.2$ (blue solid), to the $c = 20.4$ case (black solid), for internal clump densities set to the Hernquist and Jaffe profiles.}
   \label{fig:comparhalodens}
\end{figure*}

As may be expected from the above discussion relating to the formation of the isothermal core, the associated effect of density core formation is also larger when the halo is less concentrated (and so its particles less bound). This is confirmed in the plots shown in Fig.~\ref{fig:comparhalodens}. (It is worth noting, however, that because the initial halo central density is also smaller in this case, the dynamical friction coupling is actually smaller, and so the effect is initially slower than in the $c=20.4$.
case.) 

\subsubsection{Total density profiles}

\begin{figure*}
	\includegraphics[width= 8.5cm ]{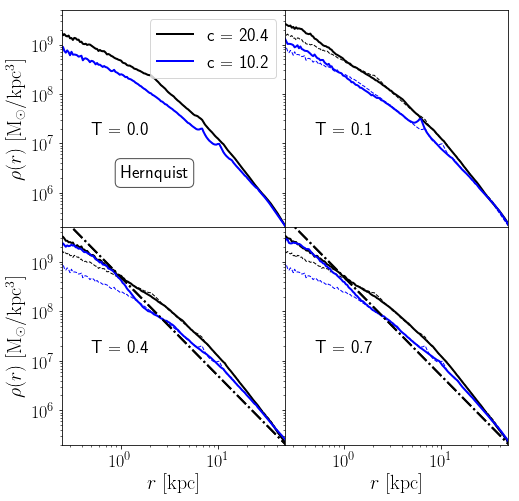}
	\includegraphics[width= 8.5cm]{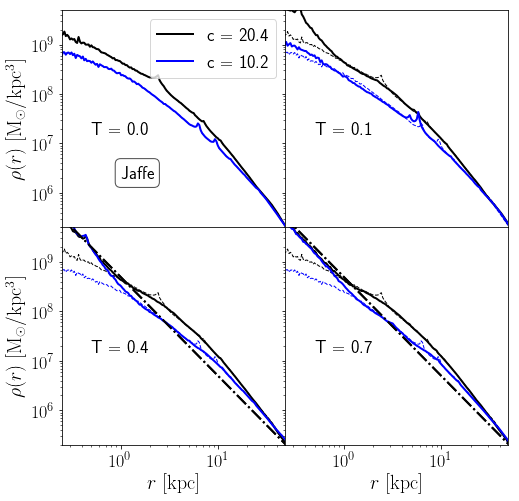}
   \caption{Total (Halo+Disc+Clumps) density profile  for halo concentrations $c=20.4$ (black solid) and $c = 10.2$ (blue solid), all other parameters remaining fixed. The dashed lines on late time plots correspond to singular isothermal ($1/r^2$) profiles.}
\label{fig:totdens}
\end{figure*}

The cusp-core transformation is associated with a 
decrease in halo density, and hence in dark matter mass enclosed within a 
given radius in the region where the cusp reigned. 
Nevertheless, despite the decrease in central halo mass, 
the {\it total} mass in the central 
region actually increases, as the central bulge, 
born of clump remnants, materialises.
This result, which we illustrate in Fig.~\ref{fig:totdens},
is in contrast to the case when clumps are distributed  initially in the same way as the smooth background (NFW) halo; in this case the total density profile remains invariant to a good approximation (\citealp{EZetal2004,EZ2008, EZJPhC2019}). It is also opposite to the case when the cusp-core transformation is due to fluctuating feedback driven gas; in that case the total density decreases and the core forms due to the shallowing of the central potential 
(\citealp{Haloheatin2023}; Appendix~B).

When the effect of the dynamical friction is strongest --- namely in the case
of the Jaffe internal clump profile --- the total inner density distribution in fact converges towards a logarithmic slope $\sim -2$, after sufficient time. This is in line 
with observations of the central parts of early type galaxies, where an inner 'universal' profile is found,  and  leads to interesting consequences regarding the 
rotation curves of spiral galaxies as well, as we describe below.
(These issues are also discussed further in the concluding section).

\subsection{Rotation curves: diversity and bulge halo conspiracies}

\begin{figure*}
	\includegraphics[width= 17cm ]{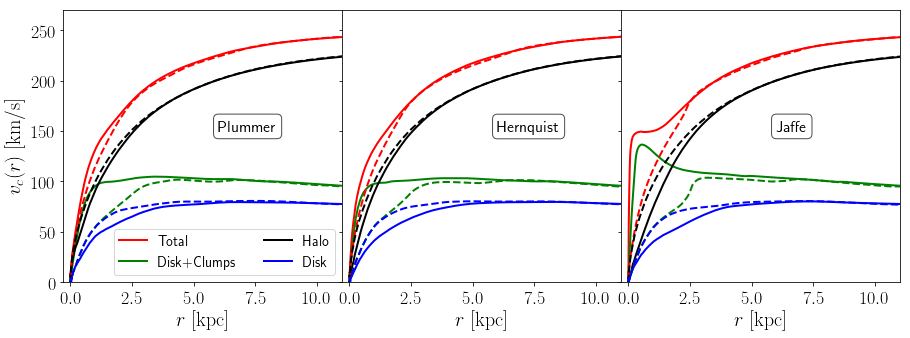}
    \caption{Rotation curve of dark halo, stellar disc and clumps, as well as the total, for all simulations  with fiducial disc-halo parameters (Table~\ref{table:hostprop}), shown at time $ T = 0.7~{\rm Gyr}$ (solid lines), and initially at $T = 0.0$ (dashed lines).} 
    \label{fig:RC}
\end{figure*}

\begin{figure*}
	\includegraphics[width= 7.5 cm]{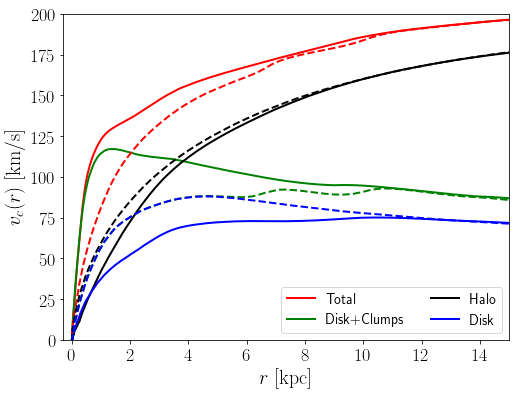}
 \hspace{1cm}
	\includegraphics[width= 7.5 cm]{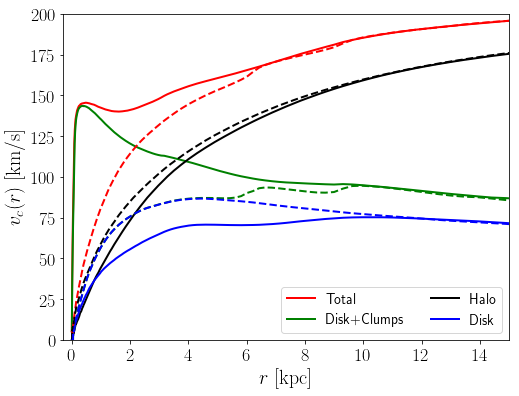}
   \caption{Initial ($T=0$) and final ($T= 0.7 {\rm Gyr}$) rotation curves in simulations with Hernquist (left) and Jaffe-profile clumps with halo concentrations $c=10.2$.}
\label{fig:RC_c10}
\end{figure*}

\begin{figure}
	\includegraphics[width= 8cm]{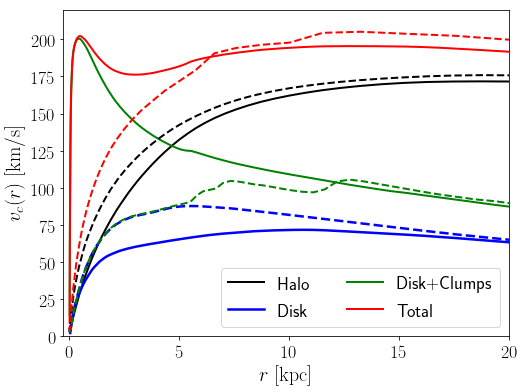}
    \caption{Initial ($T=0$) and final ($T= 0.7 {\rm Gyr}$)  rotation curves in simulation with Jaffe-profile clumps and halo  concentration $c=20.4$, but with  halo mass reduced by half and stellar mass (including both disc and clumps) doubled.}
    \label{fig:conspiracy}
\end{figure}

\begin{figure*}
	\includegraphics[width= 16 cm, height= 6.6 cm]{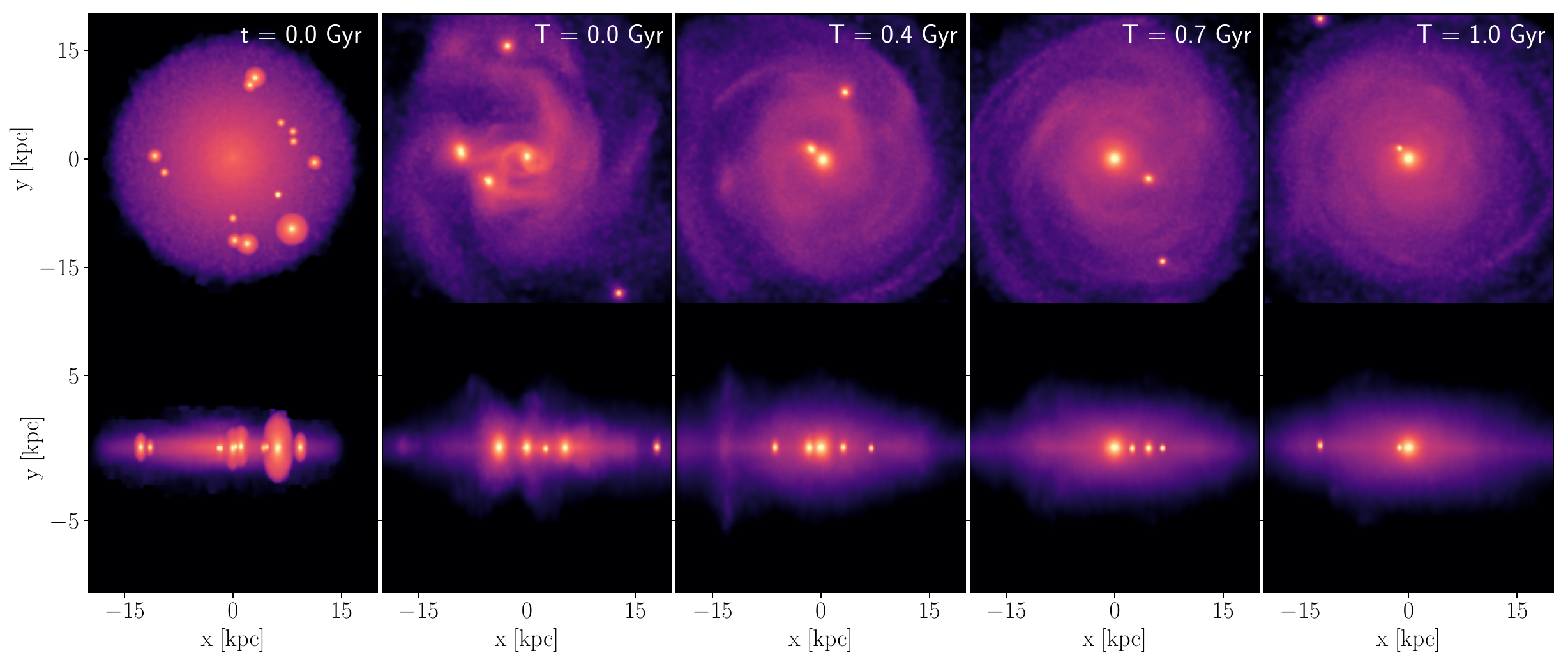}
    \caption{Same as Fig.~\ref{fig:DiskClumpImg}, with Jaffe profile clumps, and  when the halo mass  is halved halo and stellar mass (disc plus clumps) is doubled.}
    \label{fig:OptSIMProjectedDensity}
\end{figure*}

In Fig.~\ref{fig:RC}, we plot the rotation curve (defined simply in terms of the mass enclosed within a given radius) of the halo, disc and  clumps, as well as  the total, for simulations of our fiducial disc-halo galaxy model with embedded clumps endowed with the different internal  profiles considered.
Initially, for the parameters chosen, 
the contribution of the halo is largest even 
towards the centre of our simulated galaxy, which 
does not initially contain a bulge. As  a result of the 
clump migration, however, a central  bulge forms; a process that is accompanied
by decrease in the central dark mass, as probed in previous subsections. The net result is that, starting from dark matter domination in the central region, one arrives at baryon domination in that region (as is likely the case in bright spiral 
galaxies such as the Milky Way). The evolved rotation curve now shows a central feature 
reflective of the nascent bulge. Furthermore, as may be expected 
from the previous subsections, the effect of decrease of 
central dark matter mass and increase in inner baryonic mass is enhanced as the internal clump density profiles become steeper, as one moves from the cored Plummer sphere to a singular isothermal 
centre of a Jaffe model.

The effect of halo mass ejection from the central region, 
and replacement with a
baryonic bulge, is also more significant when a 
less concentrated initial halo is used, as can be seen 
from Fig.~\ref{fig:RC_c10}. Here, the central baryonic and dark matter contribution is comparable at the effective start of the simulation at $T=0$, instead of the halo entirely dominating at all radii. But the baryon domination is complete at the end of the runs. This also comes with the emergence of 'bulge-halo conspiracy', whereby 
the two components' contributions approximately match at some transition radius to give a flat rotation curves, despite one component being dominant in the central kpc or so, and the other contributing most outside that radius.

We also ran  simulations with 
 parameters that further clarify the possible origin of such a bulge halo conspiracy,  including the advent of  a central bump of the type often observed 
in galaxies. For this purpose, 
we start from a stellar component (disk plus clumps) that is doubled in mass (in both the disc and clump masses) and a halo mass that is halved (while fixing $ c = 20.4$). The resulting rotation curves are shown in Fig.~ \ref{fig:conspiracy}, for Jaffe profile clumps. 
The final configuration --- with central bump 
linking the transition from baryonic to dark matter domination, resulting in a   
nearly flat rotation curve despite the transition  ---
is akin to that of many observed galaxies, including that of the 
Milky Way (e.g. \citealp{Rotcurverev2017PASJ}). 
The corresponding projected density plots, shown in Fig.~\ref{fig:OptSIMProjectedDensity} also show a thickening of the disc during the evolution, resulting from the interaction with the clumps, which finally spiral in and merge to form a prominent central bulge.  

\section{Conclusion}
\label{sec:conc}

We performed simulations of model disc galaxies with 
massive stellar clumps embedded in the discs. The effective 
initial conditions were designed so that the initial masses 
and sizes of the clumps are compatible with those observed 
in the HST Dynamo sample of local galaxies. 

The clumps are coupled through dynamical friction to the host disc-halo system, and so spiral to the centre (within a few hundred Myr to $\la {\rm Gyr}$), forming a central bulge. In the process, the central halo is dynamically heated; with the 'temperature inversion', 
characteristic of the initial density cusp, giving way to an isothermal core. The combination of centre-bound clump migration and outward flow of dark matter leads to the baryonic component dominating in the inner region, even when starting with a model galaxy that is dark matter dominated at all radii. 

The efficiency of the aforementioned  processes depend on the rate at which clumps are stripped as they move the disc-halo potential. which in turn depends on the form of their internal density profiles. We tried three different forms. In order of less to more steep inner density radial variation, these were the Plummer profile (with inner core), the Hernquist profile (with $1/r$ cusp), and the Jaffe profile (with $1/r^2$ cusp). We find that the steeper the inner clump profile --- and so, the more mass in the inner region, given the same total mass and tidal radius ---  the more effective the process of clump migration and concurrent effect of halo cusp-core transformation.

The replacement of the inner dark matter mass by the baryonic component of clump remnants thus effectively depends on 
the competition between two effects, namely dynamical friction and stripping.  Given that the efficiency of the stripping also depends on the host halo mass and concentration, being more efficacious when these are increased, more prominent bulges and larger halo cores are thus naturally produced if the halo concentration is decreased,  or when the halo mass is decreased relative to the disc-clump system (in the latter case the effect being particularly prominent due to larger clump masses). 

In this context, the variation of the inner clump profiles, as well as the halo concentration and mass of the halo relative to the stellar component, lead to the advent of a diversity in final rotation curves; from minor inner hump, reflective of the transformation from dark matter to baryon dominance in the very inner region, to the building of a more prominent bulge (Fig.~\ref{fig:RC}, \ref{fig:RC_c10}). 
When the effect is strongest, this bulge 'conspires' with the nascent cored halo component to produce produce a flat rotation curve, complete with the characteristic inner bump heralding the transition from baryon to dark matter dominance (\ref{fig:conspiracy}). 

Such conspiracies, as reflected in the rotation curves of disc galaxies, were once thought to characterise most  
such galaxies. But already since the 1990's the consensus shifted towards the realisation that there was a diversity of rotation curves, and that the conspiracy was only present to high surface brightness galaxies (\citealp{1992PASP..104.1109A}). More recent research has been concerned with the bulge halo conspiracy in early type galaxies, particularly with the seemingly universal isothermal total (baryonic plus dark) density profile characterising of such systems (e.g.~\citealp{Koopmans_2006, Gavazzi_2007, Duttreu2014MNRAS.438.3594D, etalkravts2014MNRAS.437.3670C, 2017ApJ...840...34S}). 

This isothermal profile, which also seems to characterise the total mass distribution of bright disc galaxies over a large radial span (\citealp{Dichotomy2019}), is reproduced as the final state in our simulations; particularly those with parameters chosen such that the dynamical friction coupling is most efficient and stripping least effective (Fig.~\ref{fig:totdens}). On the other hand, 
the variety of end states we find, depending on the relative importance of these two effects, may provide further insight, which may be added to existing explanations and interpretations of the diversity problem of spiral galaxy rotation curves (e.g., \citealp{Divers22022JCAP}).

Thus, to sum up. The migration of  the stellar clumps 
leads to the formation of halo cores that replace the initial inner dark matter density cusps with a baryonic central bulge. Depending on the inner profile of the clumps, the halo concentration, and its mass relative to the stellar component, a variety of final mass distributions may arise, including those associated with a singular isothermal total mass distribution characteristic of massive disc and early type systems. The  variety of possible end states may also form a contributing factor to the issue of diversity of rotation curves in spiral galaxies and total mass distribution of early type galaxies. Indeed, in the present scenario,  the steepening of the inner 
of the inner profile may be expected to correlate with surface brightness (e.g., with a massive stellar component and less concentrated halo) as  observed, a contention that may be tested in detail with future simulations. 

In concluding,  
it is important to note that our purely N-body simulations of the observed stellar clumps naturally do not include feedback, from either starburst or AGN, which may reduce the inner baryonic dominance that is the final result of our simulations, and perhaps even lead to total reduction in central mass distribution, contributing further to the diversity of observed rotation curves. Another related issue of course pertains to the survival of the observed stellar clumps, at least on timescales of the order of disc rotation period at solar radius in a Milky Way like galaxy. In the present study we have assumed that they may be disrupted through stripping but not through internal starburst in a major gaseous component. This is suggested by the observations in~~\cite{2022MNRAS.512.3079A},  which suggest that
the clump sample on which this study is based is relatively robust
against such effects.  They are also 
supported by simulations showing that the relative preponderance of the most massive stellar clumps, which  are primarily behind the processes described here, may actually increase as a result of feedback (\citealp{Faure_Bournaud2021}).  Further progress in examining these issues, as well as more information regarding the internal clump profile, which is our main unknown here, may be provided by future  observations in tandem with detailed simulations.

\section*{Acknowledgements}
We would like to thank Sarah Sweet for providing rotation curve data of DYNAMO-HST galaxies used in this paper and the referee, Fr\'ed\'eric Bournaud, for a constructive report.  
AAEZ would like to thank Fran\c{c}oise Combes for useful discussions. 

\section*{Data Availability}
Data is available upon request.


\bibliographystyle{mnras}
\bibliography{dynamo_clumps_galsim}



\appendix
\section{Host Equilibrium Check and convergence test}
\label{app:conv}
\begin{figure}
	\includegraphics[width=\linewidth]{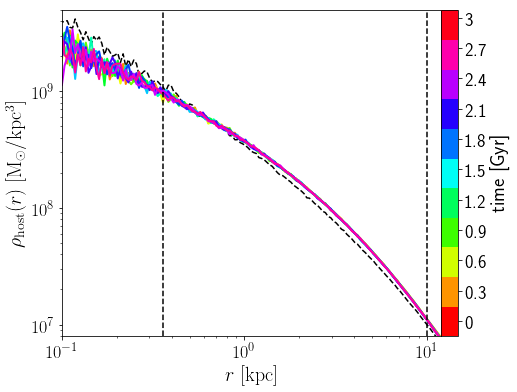}
    \caption{The dark matter halo density profile over 3 Gyr of evolution time, in a system with disc and halo (without massive clumps). The two vertical dashed lines denote the convergence radius at the maximum simulation time of 3 Gyr, $r_{\rm cov} = 0.36\; \rm kpc$ (cf. equation~\ref{eq:rcov}), and NFW the scale radius $r_s$. Note that the results shown in this work concern a period of effective evolution (after interval of relaxation to quasi-equilibrium) of 0.7~Gyr, and thus the relevant convergence radius is smaller than shown here.}
    \label{fig:HostDensProf}
\end{figure}

We seek to verify that our disc-halo configurations do not evolve 
significantly in the absence of the stellar clumps;  that they quickly reach a quasi-steady state on a timescale much smaller
than that associated with the effect of dynamical friction mediation between the clumps and the host system, and that they remain 
in that state. 
In Fig.~\ref{fig:HostDensProf} we show results 
of a simulation that is run with disc and halo only, without the 
added stellar clumps. The convergence radius $r_{\rm cov}$ is defined such that the maximum simulation time ($t = 3\; \rm Gyr$) is equal to the two-body relation time $t_{\rm relax}$ which is given by (\citealp{2003MNRAS.338...14P}),
\begin{equation}
    t_{\rm relax}(r) = \frac{\pi N(<r)}{4\ln N(<r)} \sqrt{\frac{r^3}{GM(<r)}},
    \label{eq:rcov}
\end{equation}
where $N$ and $M$ are the host particles number and mass within radius $r$. As the results shown in the rest of this work concern a 
period of less than 1 Gyr, the the relevant convergence radius 
is smaller still. 
Also, as can be seen, except for initial relaxation 
lasting about 30 Myr (much less than a typical clump orbital period), the dynamical structure of the host system remains stable.

\begin{figure}
	\includegraphics[width=\linewidth]{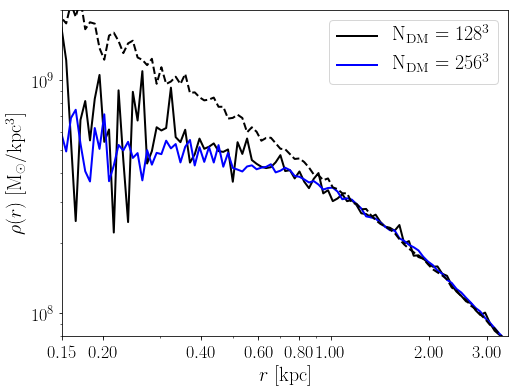}
    \caption{Dark matter halo density profile in simulation with the same host and clump properties but different mass resolutions. The results are shown at $t = 1.2\; {\rm Gyr}$, with the dashed-black line showing the density distribution at $t = 0.3\; {\rm Gyr}$. The internal clump profile is set to the Jaffe model and the disc and halo parameters correspond to the fiducial runs of Table~\ref{table:hostprop}.}
    \label{fig:ClumpDensCompRes}
\end{figure}

We have also verified that our results are insensitive 
to the particle numbers used in the simulation. Fig.~\ref{fig:ClumpDensCompRes} shows an example of such a control run.

\section{Truncated clump mass as function of orbital radius}
\label{app:Jacoby}
\begin{figure}
	\includegraphics[width=\linewidth]{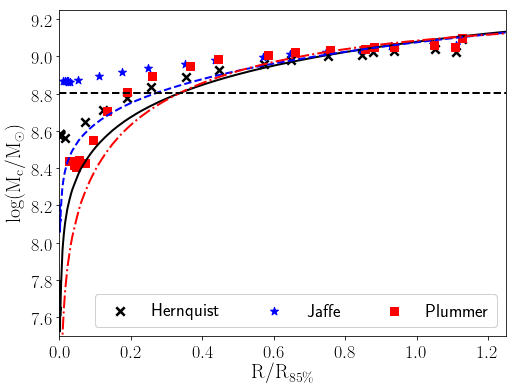}
    \caption{Mass versus orbital position of a one clump simulations with mass $2\times10^{8.5} M_{\odot}$, overplotted with estimations from the Jacobi formula (eq. \ref{eq:jacobiRd}). The horizontal black-dashed line is at $2 \times 10^{8.5} M_\odot$.} 
    \label{fig:ClumpPosEvol}
\end{figure}

In order to examine mass stripping of clumps due to tidal forces in a more controlled manner, and compare with estimates using the Jacobi radius, we ran one-clump simulations. An example is shown here, where the clump is set initially at orbital radius $R = 11.9 {\rm kpc}$, and has mass $M_c=  4\times 10^{8.5} M_\odot$. In Fig.~\ref{fig:ClumpPosEvol}, the evolution of the clump mass as a function of orbital radius is plotted together with theoretical estimates using the Jacobi formula (equation \ref{eq:jacobiRd}).  
The extent of mass preservation depends on the internal clump mass profile. 
But it is only sensitive to the particular profile in the inner region. 
This enables the advent of  effective initial conditions, after 
about half a Gyr relaxation period, that are similar for all internal 
clump mass distribution used in this work (as described in Section~\ref{sec:ClumICs}). 
Furthermore, the Jacobi estimates of the mass stripping  agree relatively well with the simulation results at larger orbital radii and fail at smaller ones. This also reinforces the plausibility of the procedure used in Section~\ref{sec:ClumICs}, which involves truncating the clumps at the Jacobi radius
at large radial distances $R$ in the disc and letting the system of clumps relax, to the approximate the observed distribution, from there. In the process, the idealised initial spherical clump configurations are also distorted as they adopt to the host potential.


\bsp	
\label{lastpage}
\end{document}